# Pressure-induced anomalies in the magnetic transitions of the exotic multiferroic material, $Tb_2BaNiO_5$

K.K. Iyer,[1] Ram Kumar,[1] S. Rayaprol,[2] K. Maiti[1] and E. V. Sampathkumaran[2,3]

[1]*Department of Condensed Matter Physics and Materials Science, Tata Institute of Fundamental Research, Homi Bhabha Road, Colaba, Mumbai 400005, India*
[2]*UGC-DAE-Consortium for Scientific Research, Mumbai Centre, BARC Campus, Trombay, Mumbai 400085, India*
[3]*Homi Bhabha Centre for Science Education, Tata Institute of Fundamental Research, V. N. Purav Marg, Mankhurd, Mumbai, 400088 India*

We have studied the influence of external pressure up to 1 GPa on the magnetic transitions of the orthorhombic Haldane-spin chain compound $Tb_2BaNiO_5$, an exotic multiferroic material. This parent compound is known to undergo Néel ordering at $T_{N1}$= 63 K and another magnetic transition at $T_{N2}$ = 25K at which ferroelectricity sets in, however, without any change in the magnetic symmetry, but with only a sharp change in the canting angle of Tb 4f and Ni 3d magnetic moments. There is a subtle difference in the antiferromagnetic state above and below $T_{N2}$, which is supported by the fact that there is a metamagnetic transition below $T_{N2}$ only (for 5 K, at ~60 kOe). We report here that, with the application of external pressure, there is an upward shift of $T_{N1}$, while $T_{N2}$ shifts towards lower temperatures. It is interesting that the two magnetic transitions in the same compound behave differently under pressure and the opposite behavior at $T_{N2}$ is attributed to local distortion leading to ferroelectricity. The results are augmented by temperature dependent x-ray diffraction and positive chemical pressure studies. The chemical pressure caused by the isoelectronic doping at Ba site by Sr reduces both the transition temperatures. Clearly, the external pressure favors antiferromagnetic coupling (that is, leading to $T_{N1}$ enhancement), whereas the chemical pressure reduces $T_{N1}$, suggesting important role of the changes in local hybridization induced by doping on magnetism in this material.

# 1. INTRODUCTION

The investigation of the mutual influence of two historically known antagonistic phenomena, namely, magnetism and ferroelectricity in the same compound, has been one of the active topics of research in condensed matter physics. The spin-induced ferroelectric behavior, called 'type-II multiferroicity', is now commonly known in many compounds with magnetism from the transition metal ions playing a key role on the so-called 'multiferroic' phenomenon [1]. This becomes interesting because ferroelectricity historically is associated with $d^0$-ness, whereas magnetism requires unpaired electrons in the d-orbitals. The magnetic frustration arising out of certain (e.g., triangular, kagome) geometrical arrangement of antiferromagnetically coupled transition metal ions has been known to be an essential factor to induce multiferroicity in many materials. There are sufficient experimental evidences on some of the most celebrated multiferroics, e.g., $R$MnO$_3$, $R$Mn$_2$O$_5$, CuCrO$_2$, and Ni$_3$V$_2$O$_8$, (where $R$ = Rare-earth), for the influence on ferroelectricity by a change in magnetic structure induced by external parameters, for example, high pressure, $P$ [2-9]. This is due to the fact that the external pressure in general can lead to a change of the magnetic exchange coupling constants by compressing the lattice, decreasing the interatomic distances and possibly changing the bond angles between different ions, which in turn can affect the magnetically ordered phases as well as the ferroelectric displacements. Additionally, the presence of magnetic rare-earths, ordering magnetically at much lower temperatures can complicate magnetoelectric coupling and pressure dependence of ferroelectric properties further in rare-earth containing multiferroics. As a result, many of these materials, in particular rare-earth containing manganites, are characterized by a cascade of magnetic phase transitions with decreasing temperature. It was found that various interactions compete with each other under pressure, e.g., first and second nearest-neighbor isotropic Heisenberg interactions as well as magnetic anisotropy, antisymmetric Dzyaloshinskii-Moriya interactions, symmetric exchange interactions and p-d hybridization, depending on the material on hand. If one disregards rare-earth induced low-temperature magnetic transitions and focusses on the magnetic ordering of the Mn sublattice responsible for ferroelectricity, it was found that the initial antiferromagnetic ordering temperature and the lower one responsible for ferroelectricity are found to increase with pressure in the case of RMnO$_3$ [6, 7] and RMn$_2$O$_5$ [8]. In the case of the delafossite, CuCrO$_2$, the spin spiral state onset temperature (at 24 K), at which ferroelectric order occurs, increases with pressure [3]. In the kagome lattice, Ni$_3$V$_2$O$_8$, the temperature (around 6.5 K) at which helical spin density wave structure breaks the spatial inversion symmetry decreases marginally with pressure [9]. Clearly, no systematics have evolved till now with respect to pressure dependence of multiferroic temperature. Barring such reports, high pressure studies on multiferroics are less abundant and it is therefore of interest to gather knowledge about the influence of pressure on different spin-induced multiferroics.

With this background, we performed high pressure studies on the Haldane spin-chain compound, Tb$_2$BaNiO$_5$ [10, 11], crystallizing in the orthorhombic structure (space group *Immm*) [12-15]. In this structure, the corner-sharing NiO$_6$ octahedra run along *a*-axis forming chains (isolated by Tb and Ba ions) and these octahedra are distorted in the sense that Ni-O apical distance is less than that in the basal plane and O-Ni-O bond angle is also reduced compared to that for regular octahedron. This compound provides a different and unique opportunity for high pressure studies with respect to the situation in the compounds mentioned above, as elaborated here. This compound has been recently reported to be exotic in its magnetic, multiferroic and magnetodielectric (MDE) coupling properties [16-20]. This insulating compound exhibits two antiferromagnetic transitions [16], one at about $T_{N1}$ = 63 K and the other at about $T_{N2}$ =25 K, and, unlike in manganites mentioned above, the onset of magnetic order for both Tb-4f and Ni-3d magnetic moments occurs at the same temperature [12]. The magnetic structure is made up of mutually canted collinear 3$d$(Ni) and collinear 4$f$(Tb) sublattices even across $T_{N2}$ with temperature ($T$) dependent canting angles with respect to $c$-axis ($\theta_{Tb}$, and $\theta_{Ni}$) as well as relative canting angle ($\Delta\theta = |\theta_{Tb} - \theta_{Ni}|$). Otherwise, there is no change in the magnetic symmetry down to 2 K. However, multiferroicity is observed below $T_{N2}$ only. The fact that there is a subtle difference in the magnetic states above and below $T_{N2}$ is revealed by the observation of a metamagnetic transition below this temperature only. It was established [17] that there is a sudden increase in the canting angle of Ni and Tb magnetic sublattices at 25 K at which spontaneous electric polarization sets in with a lowering of temperature. Crystallographic features do not provide any evidence for geometrically frustrated magnetism unlike in the compounds discussed above and it appears that a new theory based on exchange striction coupled with critical canting angle may be required to explain multiferroicity in this compound [17]. In support of this, a small doping (10 atomic percent) of Sr for Ba destroys ferroelectricity and the mutual canting angle of the two magnetic sublattices are below the critical value [20]. Thus, multiferroicity in this compound is conceptually different from other compounds, providing a new landscape in the field of multiferroicity.

It may also be added that the magnetism of this compound is special within this rare-earth family, $R_2BaNiO_5$ [21], as the onset of antiferromagnetic ordering temperature, $T_{N1}$, is the largest within this family, attributable to the dominant role of single-ion 4$f$ anisotropy as stated in [16]. A strong MDE coupling is reported below $T_{N2}$ (as large as about 18%) which is unusual for a polycrystalline material [16]. Subsequent studies showed that this can be further enhanced by a small Y doping [18] and also superseded by the Co analogue [22]. This compound also presents a rare situation in which the rare-earth plays a crucial role to induce ferroelectricity [18] - unlike many other oxides in which the magnetism of transition metal ions is responsible for this phenomenon. Naturally, this compound offers an opportunity to understand the role of spin-orbit coupling [23] on ferroelectricity.

We therefore consider it important to understand magnetism of this exotic compound under pressure. With this motivation, we have carried out dc magnetic susceptibility ($\chi$) studies up to 1 GPa down to 2 K on $Tb_2BaNiO_5$. To augment the line of arguments, we present the results of our investigation on the influence of chemical pressure, induced by small Sr doping, that is, in the series $Tb_2Ba_{2-x}Sr_xNiO_5$, and of $T$-dependent x-ray diffraction (XRD) patterns.

## II. EXPERIMENTAL DETAILS

The parent compound as well as Sr-doped specimens, $Tb_2Ba_{1-x}Sr_xNiO_5$ ($x$ = 0, 0.025, 0.05, 0.075, and 0.1), were prepared in the polycrystalline form using the solid-state reaction route starting from stoichiometric amounts of high purity oxides, $Tb_2(CO_3)_2 \cdot nH_2O$, NiO, $BaCO_3$, and $SrCO_3$. The samples thus prepared were found to be single phase forming in the orthorhombic *Immm* space group by Rietveld fitting of the XRD (Cu K$_\alpha$) patterns at room temperature, similar to the ones shown in [19]. The fact that Sr goes into the lattice was further ascertained from the shift of the XRD peaks, which is found to be distinctly visible at least for $x$>0.05 at higher angle side, however small it may be. In addition, the homogeneity of the samples was confirmed by scanning electron microscope. Further characterization of $Tb_2BaNiO_5$ and $Tb_2Ba_{0.9}Sr_{0.1}NiO_5$ was carried out by $T$-dependent XRD studies down to 2 K using x-ray radiation with wavelength 0.883 Å on Indian beamline, BL-18B at High Energy Accelerator Research Organization (KEK)-Photon

Factory, Japan. The beamline energy was $E = 14.02$ keV, and it was calibrated using $LaB_6$ standard sample data. The *dc* magnetization (*M*) measurements were performed as a function of both magnetic-field (*H*) and temperature with the help of a commercial (M/s. Quantum Design) superconducting quantum interference device (SQUID). The magnetism under high pressure (0.3, 0.6, and 1 GPa) was studied for the parent compound (2 – 300 K) in a hydrostatic pressure medium (daphne oil) employing a commercial pressure cell (EasyLab Technologies Ltd, U.K.). The pressure applied on the sample was calibrated by measuring the superconducting transition temperature of Sn in the low temperature range.

## III. RESULTS AND DISCUSSIONS

The results of magnetization (*H*= 5 kOe and 100 Oe) measurements under external pressure are shown in Fig. 1a below 100 K and the curves under external pressure are shifted for the sake of clarity, as the overlap of the curves mask the features we want to highlight. We measured with two fields to convince the readers that the trends discussed here are reliable. The features in the $\chi(T)$ were discussed at several places in the past literature. Following Curie-Weiss behavior in the paramagnetic state as known earlier [16], there is a kink due to the onset of antiferromagnetic order at $T_{N1}$ (marked by a vertical arrow in Fig. 1a) at which heat-capacity was shown to exhibit a distinct anomaly [16]; a broad peak at a lower temperature (at ~ 40 K) appears which has been attributed to the persistence of Haldane gap due to Ni one-dimensional magnetism in the magnetically ordered state. The kink is very weak and therefore one has to carefully do the measurements under pressure to track its behavior. It can be inferred from the $\chi(T)$ plots as well as from the derivative plots (Figs. 1b and 1c) that $T_{N1}$ tends to increase gradually with *P*, e.g., by about 2 K for an increase of *P* to 1 GPa. However, it is not straightforward to infer $T_{N2}$ from the $\chi(T)$ plot, as the fall in $\chi$ below the peak due to the magnetic gap is rather steep and, therefore, the derivative curves are used to infer the trend. The derivative curves, $d\chi/dT$, shown in the figures 1b and 1c reveal a sudden increase in slope near 25 K under ambient pressure conditions, but the fact remains that there is a marginal downward shift of the curve around this temperature, however small it may be, with increasing pressure (by about 0.5 K for *P*= 1 GPa). Clearly, the external pressure acts in the opposite way at these two magnetic features. Finally, the effective magnetic moment ($\mu_{eff}$ = 9.63 $\mu_B$/Tb) obtained from the Curie-Weiss region (say, above 100 K) is found to be insensitive to pressure within experimental error (<0.05 $\mu_B$) and the value is in good agreement with that for trivalent Tb ion (9.72 $\mu_B$); the paramagnetic Curie temperature ($\theta_p$ = -20 ± 1K) is also found to be independent of pressure (see inset of figure 1a, comparing inverse $\chi$ plots in the high-*T* region for 0 and 1 GPa). It is also obvious from the figure 1 that there is no other change in the features, thereby establishing that the Haldane gap is not affected by external pressure up to 1 GPa.

We now compare results of high-pressure measurements of $Tb_2BaNiO_5$ with the observations made in the experiments with $Tb_2Ba_{1-x}Sr_xNiO_5$ in which a chemical pressure was induced by the partial substitution of Ba atoms by Sr atoms (Fig. 2). The curves for doped compositions are shifted along y-axis for the sake of clarity. Though we had earlier reported magnetic, and magnetodielectric coupling properties of this Sr doped series, $Tb_2Ba_{1-x}Sr_xNiO_5$, [19] for *x*= 0, 0.1, and 0.2, we felt the need to study the magnetic properties at closer intervals of *x* below $x = 0.1$ synthesized under identical conditions (that is, same batch), considering abrupt changes in the multiferroic properties for initial doping beyond $x = 0.1$. [Beyond *x*= 0.2, we see sample inhomogeneities, unlike Y-doping which can replace Tb completely]. We find that ferroelectricity is destroyed for $x = 0.15$ in this batch of specimens, and not for $x = 0.10$, thereby revealing that there is a small spread in the composition at which multiferroicity vanishes depending on the batch of specimens. We restrict the discussions here for the trends in magnetic behavior only for selected compositions, $x$ = 0, 0.025, 0.05, 0.075, and 0.1. The $\chi(T)$ and the derivative curves below 100 K, measured with 5 kOe are shown in Fig. 2. It is clear that the feature due to the transition at $T_{N1}$ shifts downwards gradually with increasing $x$, visible even for a doping as small as $x$ = 0.025. There is an overall decrease by about 5 K for $x$ = 0.1 in this investigation. $T_{N1}$ reported for $x$ = 0.2 [15] also follows this trend. As shown earlier [19], all the three lattice parameters, $a$, $b$, and $c$, undergo a weak reduction (in the third decimal place only) and the overall reduction in the unit-cell volume as $x$ is varied from 0 to 0.1 is 0.6 Å$^3$ (~ 0.24%). Since the bulk modulus of the parent compound is not known, it is difficult to quantify the pressure exerted by this change in the unit-cell volume. Nevertheless, it is straightforward to conclude that, qualitatively speaking, chemical pressure acts differently with respect to that caused by external pressure as far as $T_{N1}$ is concerned. Now, with respect to the behavior of $T_{N2}$, though it is not straightforward to infer this characteristic temperature precisely, one gets an idea by looking at the slope change of the derivative curves (Fig. 2b) in the vicinity of 25 K. It is clear that the sudden increase in the slope at 25 K noted for the parent compound gradually shifts towards low temperature range, which is a signature of corresponding lowering of $T_{N2}$. We therefore conclude that both $T_{N1}$ and $T_{N2}$ decrease with the increasing chemical pressure, exerted on the Ba layer, in contrast to the ones seen under external pressure. This means that $T_{N1}$ is sensitive to changes in local hybridization due to doping.

It is of interest to see the trend under chemical pressure by isoelectronic substitution at the Tb site. For this purpose, the readers may see [18] for the results on Tb$_{2-x}$Y$_x$BaNiO$_5$, in which the magnitude of unit-cell volume reduction for $x$ = 0.5 is comparable to that for $x$ = 0.2 of the Sr-doped series. Therefore, one can assume that the magnitudes of internal pressure are similar for these compositions. Both $T_{N1}$ and $T_{N2}$ were found to diminish linearly with $x$, investigated up to $x$ = 1.5, interestingly with the persistence of magnetically coupled ferroelectric order even in this dilute limit of Tb; for example, for $x$ = 0.5, $T_{N1}$ = ~55 K and $T_{N2}$ = ~20 K. Since Y substitution involves dilution of the Tb sublattice, for a comparison with Sr-doped series, it is necessary to normalize to Tb concentration (that is, to 2-$x$). The normalized values are about 73 K and 27 K respectively. Clearly, while the trend in $T_{N1}(x)$ in Y series is the same as that in high pressure, the increase of the normalized value by about 10 K compared to that for the parent compound under external pressure is quite profound. Assuming a linear variation of $T_{N1}$ with $P$, a pressure as large as about 6 GPa would be required to attain a value of 73 K. Y dilution studies clearly suggested Tb 4f plays a major role on the onset of ferroelectric ordering [18], as the transition temperatures decrease gradually with Y doping. In the crystal structure, the vertex-shared (compressed) octahedra of NiO$_6$ chains are isolated by Tb and Ba polyhedra and therefore the super-super exchange mechanism of the types Ni$^{2+}$-O$^{2-}$–R$^{3+}$-O$^{2-}$–Ni$^{2+}$ and Ni$^{2+}$-O$^{2-}$–Ba$^{2+}$-O$^{2-}$–Ni$^{2+}$ are known to be operative to induce magnetic ordering. As stated at the introduction, both the magnetic sublattices order at the same temperature. A dilution of R sublattice (by Y) has of course a natural destructive influence on the transition temperatures. But the fact that this dilution tends to enhance the scaled-$T_{N1}$ significantly (rather than remaining constant as expected for scaling with the concentration of magnetic ions) implies that the superexchange pathway involving R$^{3+}$ is the dominating one, getting stronger with the lattice pressure. While Ba is replaced by Sr, $T_{N1}$ does not follow the external pressure effect, and therefore we are tempted to argue that the superexchange path involving Ba counteracts the one caused by pressure. It is interesting that the behavior of normalized $T_{N2}$ is different; that is, the increasing trend with Y-induced chemical pressure in the Tb sublattice is opposite to that observed by external pressure or Sr-doping. We therefore believe that the changes in bonding strengths caused by isoelectronic substitution bear a profound effect on the properties.

We have also measured isothermal $M$ to see the influence of pressure on the metamagnetic transition field ($H_c$), occurring at 60 kOe at 2 K, for the parent compound. The results obtained under pressure are shown in figure 3a, and for Sr-doped compositions in figure 3b. Though we measured at various temperatures below 20 K, we show the curves for 5 K only. There is a weak hysteresis around $H_c$ as reproduced in the inset of figure 3a from our past work [16, 18, 19], but we restrict the plots here to the virgin curves only in the mainframe of figure 3a, that too in the vicinity of the field-induced transition for the sake of clarity. The curves around $H_c$ are broadened with increasing Sr-doping (figure 3b), clearly due to increasing chemical disorder. The point to be noted is that there is a marginal increase in $H_c$ with external $P$. However, $H_c$ gets depressed by Sr doping. It may be recalled [18] that the chemical pressure induced by Y substitution also depresses $H_c$. It is thus obvious that, as far as $H_c$ is concerned, there is no correlation of these results obtained by external pressure on the one hand and the chemical pressure (induced at R site as well as at Ba site) on the other. This endorses the conclusion made in the previous paragraph that the electronic structure changes caused isoelectronic substitutions at any site has a dominating effect on the properties.

We now offer evidence for the fact that the observed pressure effects are free from any ambiguities due to a possible change in the crystallographic symmetry. To show the absence of temperature induced changes in crystal structure down to 2 K, we measured XRD patterns with the synchrotron facility at several temperatures for $x=$ 0 and 0.1. In figure 4, we show such synchrotron-based powder XRD (S-PXRD) patterns at selected temperatures in the three temperature ranges, $T<T_{N2}$, $T_{N2}<T<T_{N1}$, and $T>T_{N1}$. The patterns, apart from establishing that the specimens are single phase (*Immm* space group) with the diffracting peaks appearing at expected angles only, reveal that there is no change in the patterns down to 2 K for both the compositions. Though there is a preferred orientation effect for a few Bragg peaks, we could index all the patterns using LeBail fit. The main point to emphasis here is that there is no change in the crystal symmetry down to 2 K. This conclusion was inferred by low temperature neutron diffraction studies as well on both the compositions [17, 20]. Since Sr substitutions exert chemical pressure without any change in the crystal symmetry, it is reasonable to assume that the external pressure of about 1 GPa also would not have caused any crystallographic change.

As mentioned earlier, there is no change in the magnetic symmetry as well down to 2 K for the parent compound, though canting angles change with varying temperature [17]. Since Sr content (measured for $x = 0.1$) also does not change magnetic symmetry [20], additional magnetic structural complexities are not expected to occur under pressure for the parent. We take this opportunity to clarify how the loss of inversion symmetry occurs in this compound – an issue remained unresolved till now. In order to understand this, we used the program FINDSYM [24] to find the magnetic space group and point group. Using the atomic positions of Tb and Ni (both magnetic ions), and assuming that moments are placed along the z-direction, the program suggests the magnetic point group to be *m'm'*2, which is a non-centrosymmetric, polar space group (*Im'm'*2), which allows polarization along the z-direction. However, the magnetic structure refinement of the neutron diffraction experimental data, shown in the Supplementary File of [17], revealed that there is a magnetic moment component along *x*-direction also in addition to *z*-component (though $M_z > M_x$), which implies canting of the moments in the *ac* plane. Therefore, considering this (that is, assigning moments along *x* and *z* directions), the magnetic point group obtained is *m'* (magnetic space group *Cm'*). This is also a non-centrosymmetric, polar space group allowing polarization in *x*- and *z*-directions. In a nutshell, for both the samples, i.e., $Tb_2BaNiO_5$ and $Tb_2Ba_{0.9}Sr_{0.1}NiO_5$, by carrying out the magnetic structural refinement across both long-range antiferromagnetic ordering temperatures considering the canting, the magnetic symmetry remains the same (*Cm'*) till the lowest temperature. In short, though canting of the Tb and Ni moments with respect to *c*-axis changes, resulting in the change of magnetic moment ($M_x$, $M_z$ and net moment *M*) as temperature is reduced, there is no change in the magnetic symmetry. This finding emphasizes the importance of an additional criterion for the multiferroic properties in $Tb_2BaNiO_5$, which we attribute to the existence of a critical canting angle.

## IV. SUMMARY

We have performed high pressure magnetization studies on an exotic multiferroic compound, $Tb_2BaNiO_5$, in which both Tb and Ni moments order at the same temperature with the involvement of Tb 4f in inducing ferroelectricity – a situation not encountered in the manganites. While the antiferromagnetic ordering onset temperature is enhanced by external pressure, the lower characteristic temperature, $T_{N2}$, at which multiferroicity has been known to set in, gets depressed by the external pressure. It is interesting that the external pressure has the opposite effect on the magnetic features at the two temperatures in the same compound. This situation is different from that noted for the Mn sublattice in the most celebrated multiferroic materials, $TbMnO_3$ [6,7] and $RMn_2O_5$ [8], including the magnetic transition temperature that is tied to the onset of ferroelectric transition. Given that both Tb and Ni order magnetically at the same temperature without any change in magnetic symmetry across $T_{N2}$, one would naively expect that both the transitions respond in the same way to the application of external pressure, in contrast to the observation. We therefore conclude that the local distortions leading to ferroelectricity influences the pressure dependence of concurrent magnetic transition temperature. We have also compared and contrasted the changes in the magnetic features with those seen by chemical pressure by isoelectronic substitution at Tb and Ba sites. Readers may recall [17] that there are prominent local lattice distortions (as reflected in O-Tb-O bond angles and Ni-O bond distances) at $T_{N2}$, where ferroelectricity also sets in, for the parent compound, which naturally get modified by chemical and external pressure as inferred from the analysis of neutron diffraction data presented in [20] for a Sr-doped specimen. There are less prominent and gradual changes in the lattice constants across $T_{N1}$ as well when *T* is lowered [see figure *S*2 in [17]] for the parent due to magnetostriction as one enters magnetically state and this also must get influenced by pressure, thereby changing hybridization. We conclude that the subtle distortions cause a profound effect on magnetic characteristics of this compound. We hope this work serves as a motivating force to extend the magnetic, optical and magnetoelectric studies by various experimental methods including neutron diffraction to much higher pressures, also in the presence of high magnetic fields, on this exotic material, as this material provides another avenue to understand various factors (including band gap changes [25]) in the evolution of the cross-coupling phenomena.

**Acknowledgements**
K.K.I. and K.M. acknowledge financial support from the Department of Atomic Energy, Govt. of India (Project Identification no. RTI4003, DAE OM no. 1303/2/2019/R&D-II/DAE/2079 dated 11.02.2020). E.V.S. thanks Department of Atomic Energy, Government of India, for awarding Raja Ramanna Fellowship. K.M. acknowledges financial assistance under the award, DAE-SRC-OI, DAE, Government of India. S.R. and K.K.I thank the Department of Science and Technology, India for the financial support, and Saha Institute of Nuclear Physics and Jawaharlal Nehru Center for Advanced Scientific Research for facilitating the experiments at the Indian Beamline, Photon Factory, KEK, Japan, under the proposal number JNC/KEK-JAP/IN-56 for synchrotron XRD measurements on Indian beamline (BL-18B) at PF, KEK, Japan. Authors also thank Gouranga Manna and Sabyasachi Karmakar for their help during the Synchrotron x-ray powder diffraction experiments. We thank Hong Jian Zhao, University of Arkansas, Fayetteville, AR, USA for his guidance to carry out symmetry analysis.

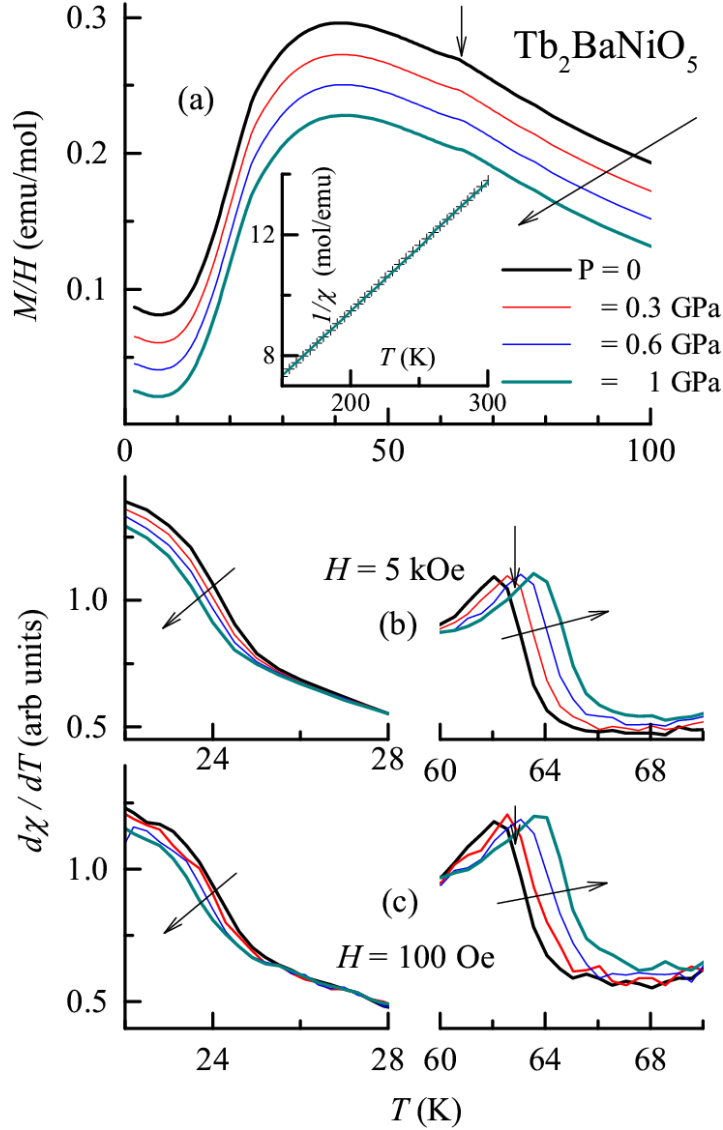

Figure 1:
Magnetic susceptibility ($\chi$) curves obtained as a function of temperature for $Tb_2BaNiO_5$ under various external pressures in the range 2-100 K measured in 5 kOe are shown in (a). The high pressure $\chi(T)$ curves are shifted for the sake of clarity, as otherwise the curves are essentially indistinguishable. Derivative curves, also for the data measured in 100 Oe, are shown in (b) and (c) in an expanded form in the vicinity of magnetic transitions ($T_{N1}$ and $T_{N2}$) described in the text. Vertical arrows mark the $T_{N1}$ and $T_{N2}$ under ambient pressure conditions. Inclined arrows are shown as guide to the eye for the reader to show the direction in which the curves move with pressure around the transition. The plot of inverse susceptibility above 150 K are shown in the inset of (a) in the absence of external pressure (+ points) and for 1 GPa (continuous line), to highlight that the paramagnetic Curie temperatures and the effect magnetic moment do not change with pressure.

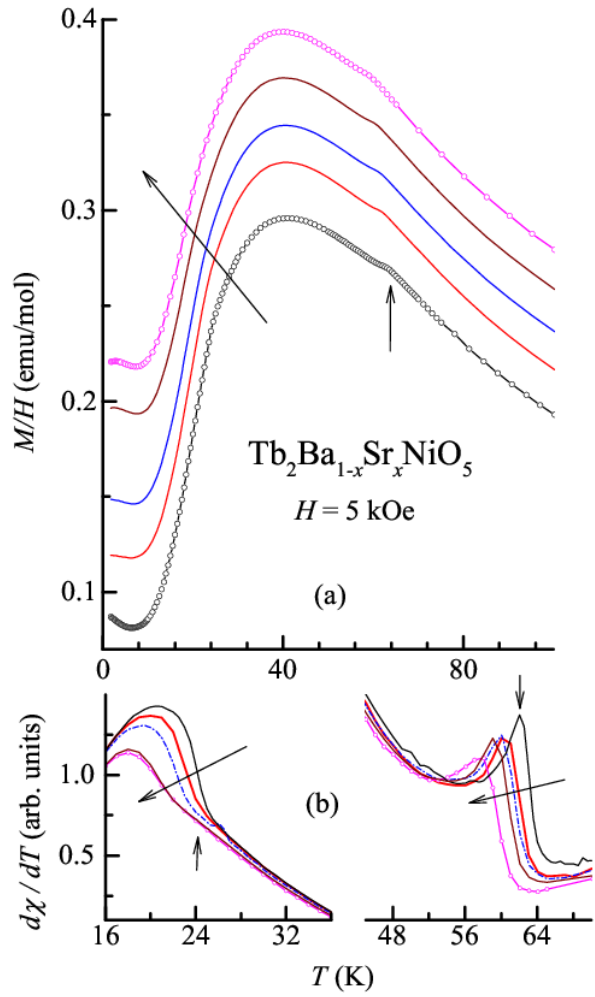

Figure 2:
(a) Magnetic susceptibility ($\chi$) as a function of temperature for Sr doped specimens, $Tb_2Ba_{1-x}Sr_xNiO_5$ in the range 2-100 K, measured with 5 kOe. In (b), the derivative curves are shown in an expanded form in the vicinity of $T_{N1}$ and $T_{N2}$. Vertical arrows mark $T_{N1}$ and $T_{N2}$. Inclined arrows are guides to the eyes to show how the transitions shift with $x$ (= 0.0, 0.025, 0.05, 0.075, and 0.1).

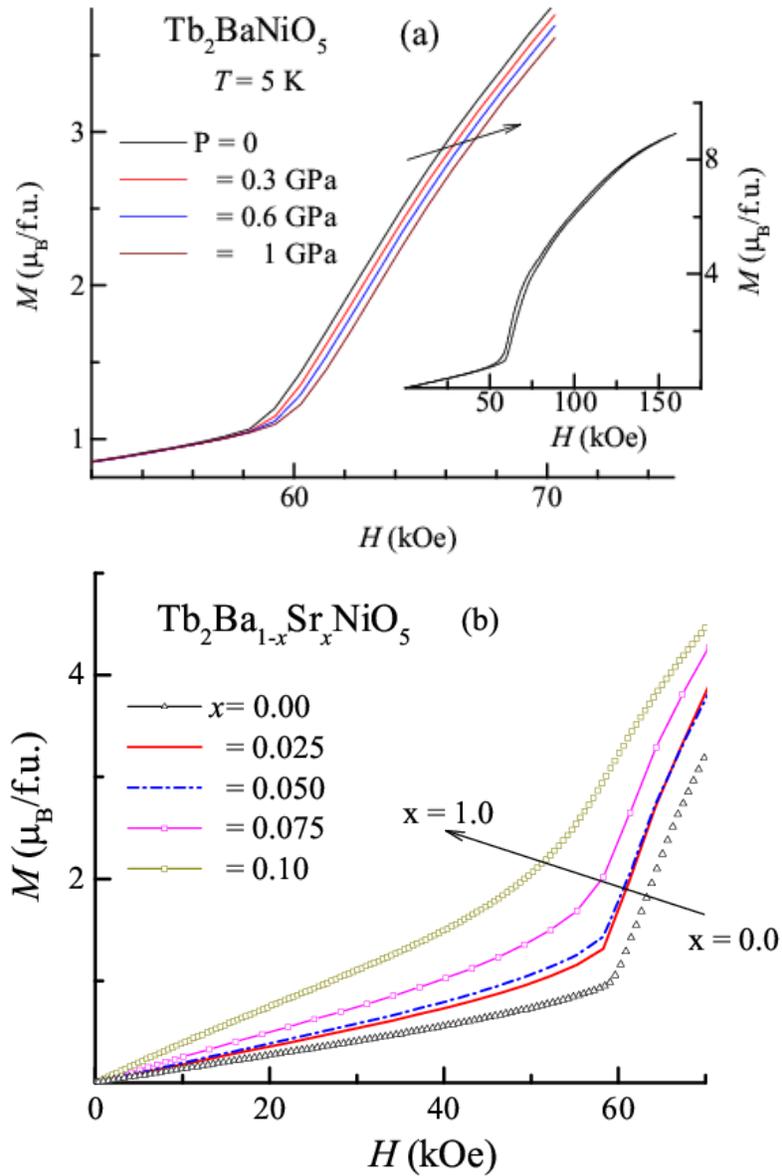

Figure 3:
(a) Isothermal magnetization (virgin) at 5 K for $Tb_2BaNiO_5$ under pressure in the metamagnetic transition region. The profile of the curve in the range 0-160 kOe under ambient pressure is shown in the inset. (b) Isothermal magnetization at 5 K for the Sr doped specimens. Inclined arrows are shown the direction in which the curves shift with increasing pressure/composition.

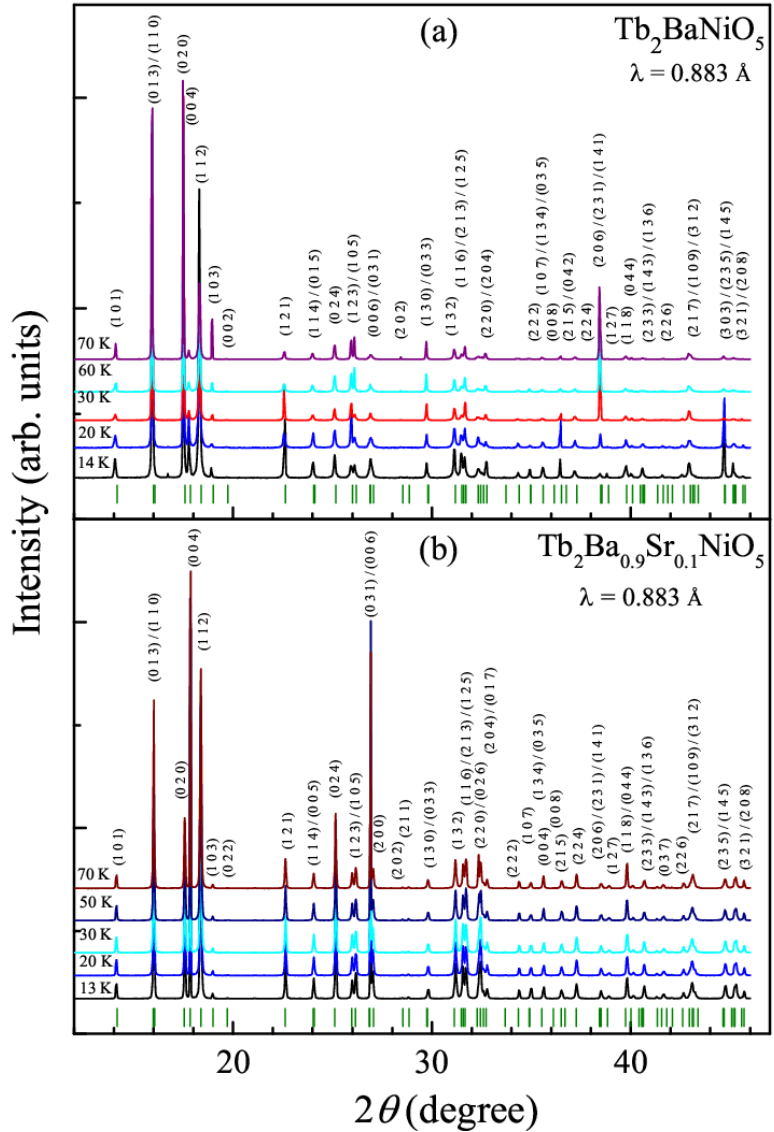

Figure 4:
Synchrotron based powder x-ray diffraction (S-PXRD) patterns of selected temperatures for $Tb_2BaNiO_5$ and $Tb_2Ba_{0.9}Sr_{0.1}NiO_5$ are shown to highlight that there is no change in structural symmetry across the magnetic transitions. Vertical bars are shown to mark the expected positions of Bragg peaks.